\documentclass{styles/svproc}
\usepackage{xcolor}

\usepackage{url}
\usepackage{siunitx}
\usepackage{graphicx}
\usepackage{placeins}
\usepackage{hyperref}
\usepackage{multicol}
\usepackage{siunitx}
\usepackage{amssymb}
\usepackage{doi}

 \DeclareSIUnit{\fos}{fps}
 \DeclareSIUnit\px{px}

\begin{document}
\mainmatter              
\title{Educational experiments with Particle Image Velocimetry in a small Prandtl water tunnel }
\titlerunning{Educational experiments with Particle Image Velocimetry}  
\author{William Thielicke\inst{1} \and Steffen Risius\inst{2}}
\authorrunning{William Thielicke and Steffen Risius} 
%
\tocauthor{William Thielicke and Steffen Risius}
\institute{Optolution Messtechnik GmbH, Gewerbestraße 18, 79539 Lörrach, Germany
\and
Kiel University of Applied Sciences (FH Kiel), Grenzstraße 3, 24149 Kiel, Germany
\email{steffen.risius@fh-kiel.de}
}
\maketitle              

\begin{abstract}
 A small water tunnel was designed as a table top experiment to visualize and measure basic fluid dynamic phenomena. The visualization of the flow by a laser light sheet and the measurement with Particle Image Velocimetry (PIV) proves to be very instructive for teaching high school and university students. Three basic fluid dynamic experiments are reported: the determination of the vortex shedding frequency of a cylinder (von-Kármán vortex street), the measurement of circulation around a flat plate and the lift coefficient of a NACA profile. All three experiments show good agreement with theoretical expectations. 
\end{abstract}
\section{Introduction}
Based on the original water tunnel setup developed by Prandtl \cite{tollmienUeberFluessigkeitsbewegungBei1961,prandtlVierAbhandlungenZur2010}, we present a new test facility for low Reynolds number flow experiments for teaching purposes. Nearly 150 years after Ludwig Prandtl's birth, his major contribution to fluid dynamics is almost unknown to high school students, and the depth of his fluid dynamic findings still poses many difficulties for university students \cite{sendAerodynamikPhysikFliegens2001}. In this article, we present an experimental test facility which allows students to understand and visualize basic fluid mechanics principles discovered more than 100 years ago using Particle Image Velocimetry (PIV) \cite{raffelParticleImageVelocimetry2018}. 
\section{Experimental setup of the water tunnel and PIV equipment}
A water tunnel (``FLOWlab''), based on the original water tunnel developed by Prandtl, was designed by FH Kiel and Optolution \cite{thielickeFLOWlabDemoEducational2023}. The water tunnel is an elongated aquarium from \SI{8}{\mm} thick glass ($1200 \times 160 \times \SI{320}{\mm}$). A horizontal black PVC wall was added at half the height of the aquarium as well as a cover for the water tunnel. This setup allows a recirculation flow inside the aquarium.
\begin{figure}[htb]
    \centering
    \includegraphics[width=1\linewidth]{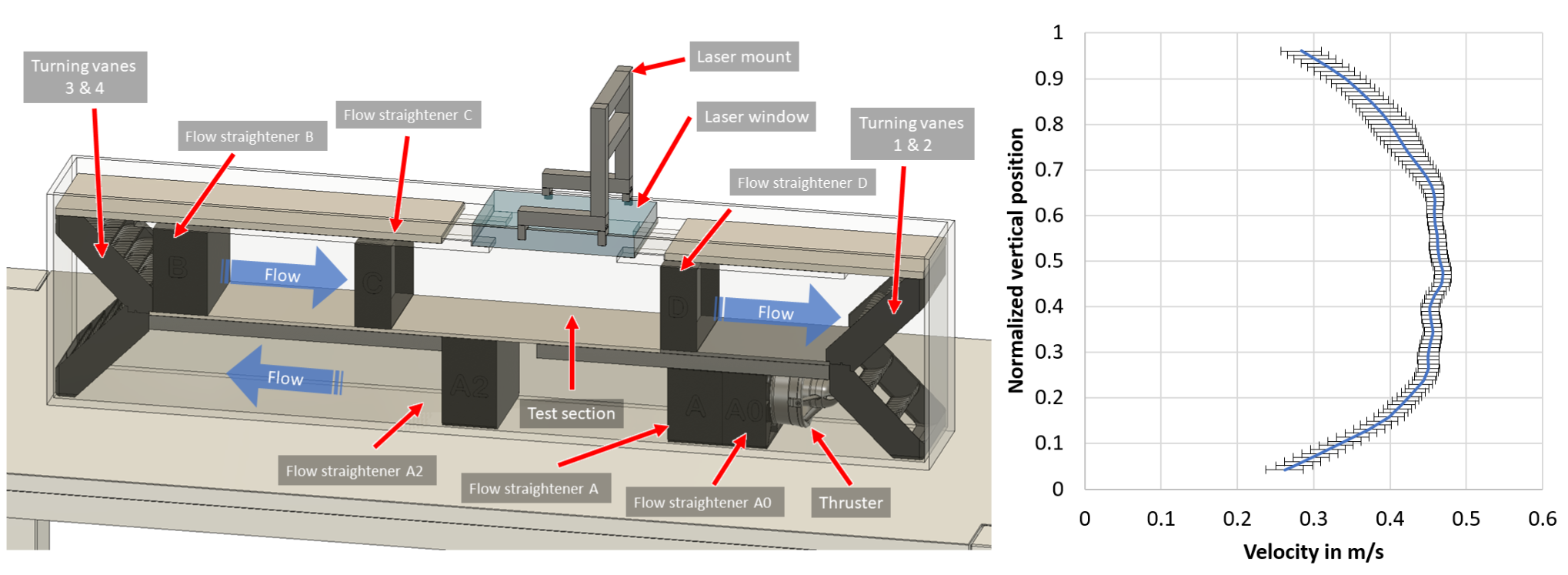}
    \caption{Left: Drawing of the compact recirculating water tunnel. The largest dimension (left to right) is \SI{1.2}{\m}. The water tunnel is filled with 60 liters of tap or distilled water. Right: Average ($n = 100$) vertical flow velocity profile in the central plane of the empty test section at maximum flow speed (\SI{0.45}{\m/\s}).}
    \label{water-tunnel}
\end{figure}
Several flow straighteners and turning vanes were added to condition the flow (Fig.~\ref{water-tunnel}). The test section measures $190 \times 125 \times \SI{140}{\mm}$. A \SI{200}{\watt} thruster, commonly used for underwater vehicles and situated in the lower tunnel section, accelerates the water in the aquarium. The thruster input voltage is reduced in the default configuration, so that a maximum flow speed of \SI{0.45}{\m/\s} is achieved. The flow is sufficiently conditioned to perform educational experiments, but due to the limited size of the whole setup, regions with lower flow velocities can be observed at the walls (Fig.~\ref{water-tunnel}).

\begin{figure}[htb]
    \centering
    \includegraphics[width=0.75\linewidth]{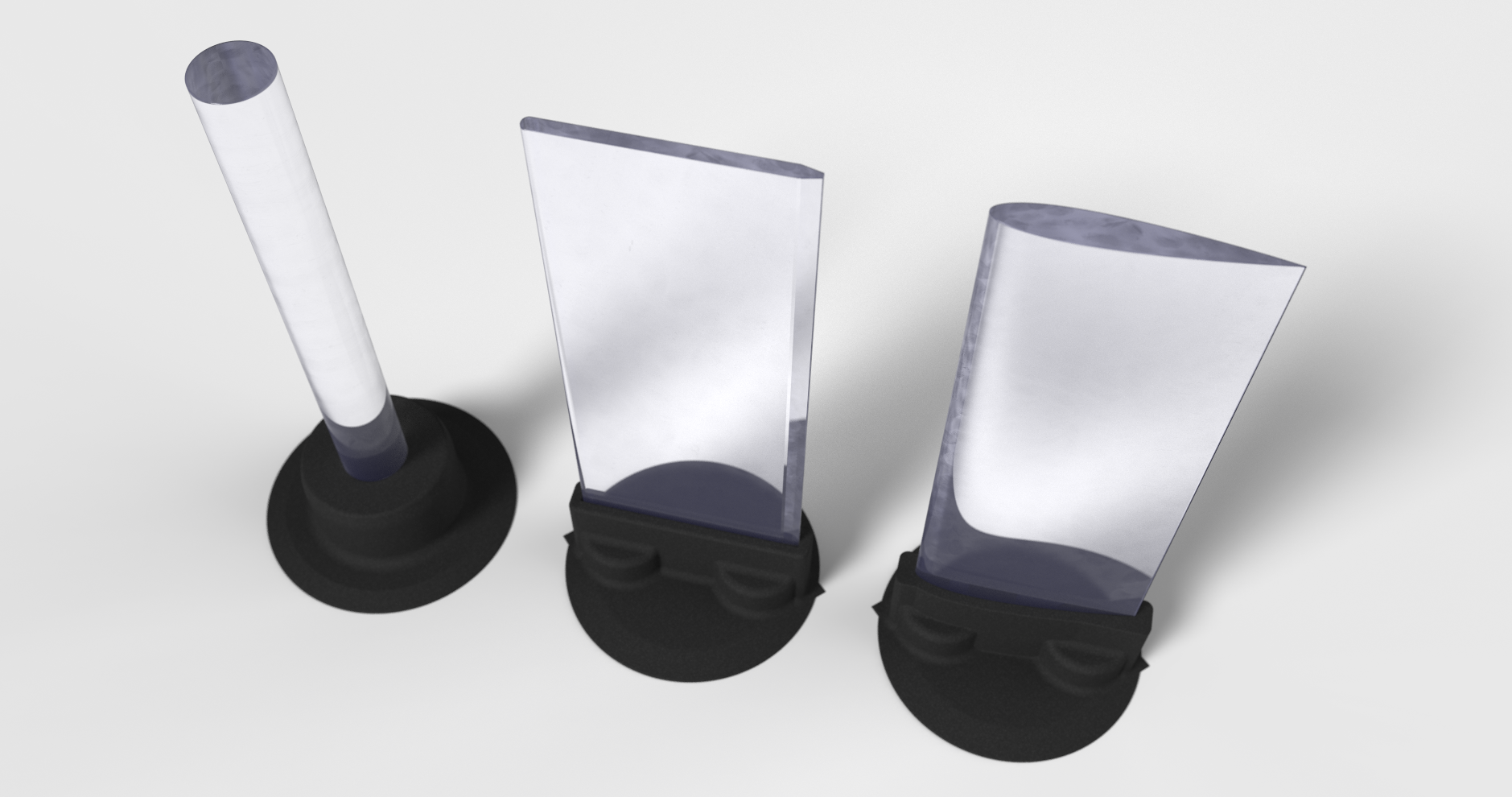}
    \caption{Flow bodies for the water tunnel. This study presents results from the cylinder (left); the flat plate (middle), and the NACA-0015 airfoil (right).}
    \label{flow_bodies_with_Naca}
\end{figure}

An Optolution PIV system was used for the experiments. Here, a low-power variant was selected, consisting of a \SI{500}{\milli\watt}, pulsed laser (LD-PS/0.5; Laser class 2M; \SI{638}{\nano\m}) with built-in sheet optics and synchronizer, allowing to use the frame-straddling technique. The integrated synchronizer is remote-controlled via PIVlab, a free MATLAB toolbox \cite{thielickeParticleImageVelocimetry2021}, and connected to a \SI{160}{fps}, $1936 \times \SI{1216}{px}$ camera (OPTOcam 2/80). The camera is equipped with a \SI{20}{\nm} wide band pass filter and a \SI{16}{\mm} lens which can also be controlled by PIVlab. The camera's field of view was set to $190\times\SI{125}{\mm}$, covering the full test section. \SI{55}{\micro\m} seeding particles from polyamide were added to the flow. The laser is attached to an optical rail on top of the water tunnel. The light sheet passes through a \SI{60}{\mm} thick perspex block that is placed on top of the water surface, avoiding any non-planar air-water interface. The camera is placed at a $90^{\circ}$ angle to the light sheet on the side of the water tunnel.
Several flow bodies can be mounted inside the test section with a magnet. The magnet outside the aquarium can be translated and tilted to reorient the flow bodies. Here, we report measurements with a \SI{15}{\mm} transparent cylinder, a \SI{3}{\mm} thick, inclined flat plate and a NACA-0015 airfoil (see Fig.~\ref{flow_bodies_with_Naca}).

The PIV analyses were performed in PIVlab \cite{thielickeParticleImageVelocimetry2021}, using multi-pass DFT cross-correlations with three passes and a final window size of \SI{24}{px} with \SI{50}{\%} overlap. For a slightly higher resolution of the flow field, we additionally used a wavelet-based Optical Flow Velocimetry (wOFV)  \cite{jassalBayesianAppproachLocally2024,schmidtImprovementsAccuracyWaveletbased2020} in some cases (using automatic smoothness detection). Standard settings for data validation and interpolation were used. The pulse distance was set between 1-\SI{2}{\ms} to achieve a tracer displacement around 5-\SI{8}{\px}. 80 double images per second were captured, 100 image pairs were analyzed for each data point.
\begin{figure}[htb]
    \centering
    \includegraphics[width=0.8\linewidth]{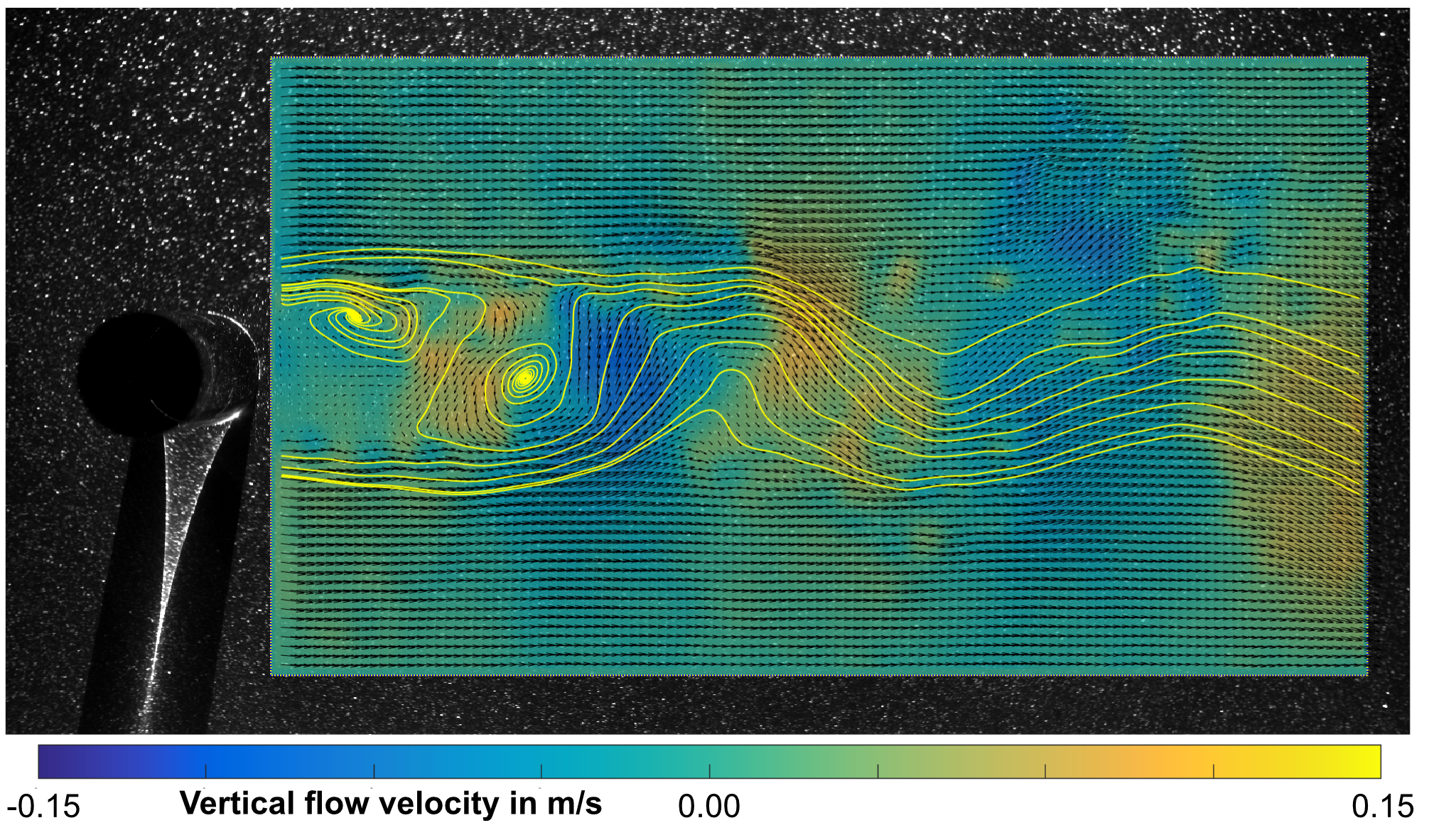}
    \caption{Instantaneous velocity field behind a \SI{15}{\mm} cylinder at $\mathit{Re} \approx 2900$. The magnitude of the velocity component perpendicular to the free stream is color coded. Streamlines derived from PIV velocity vectors are shown in yellow.}
    \label{flow_field_cylinder}
\end{figure}
\begin{figure}[htb]
    \centering
    \includegraphics[width=0.8\linewidth]{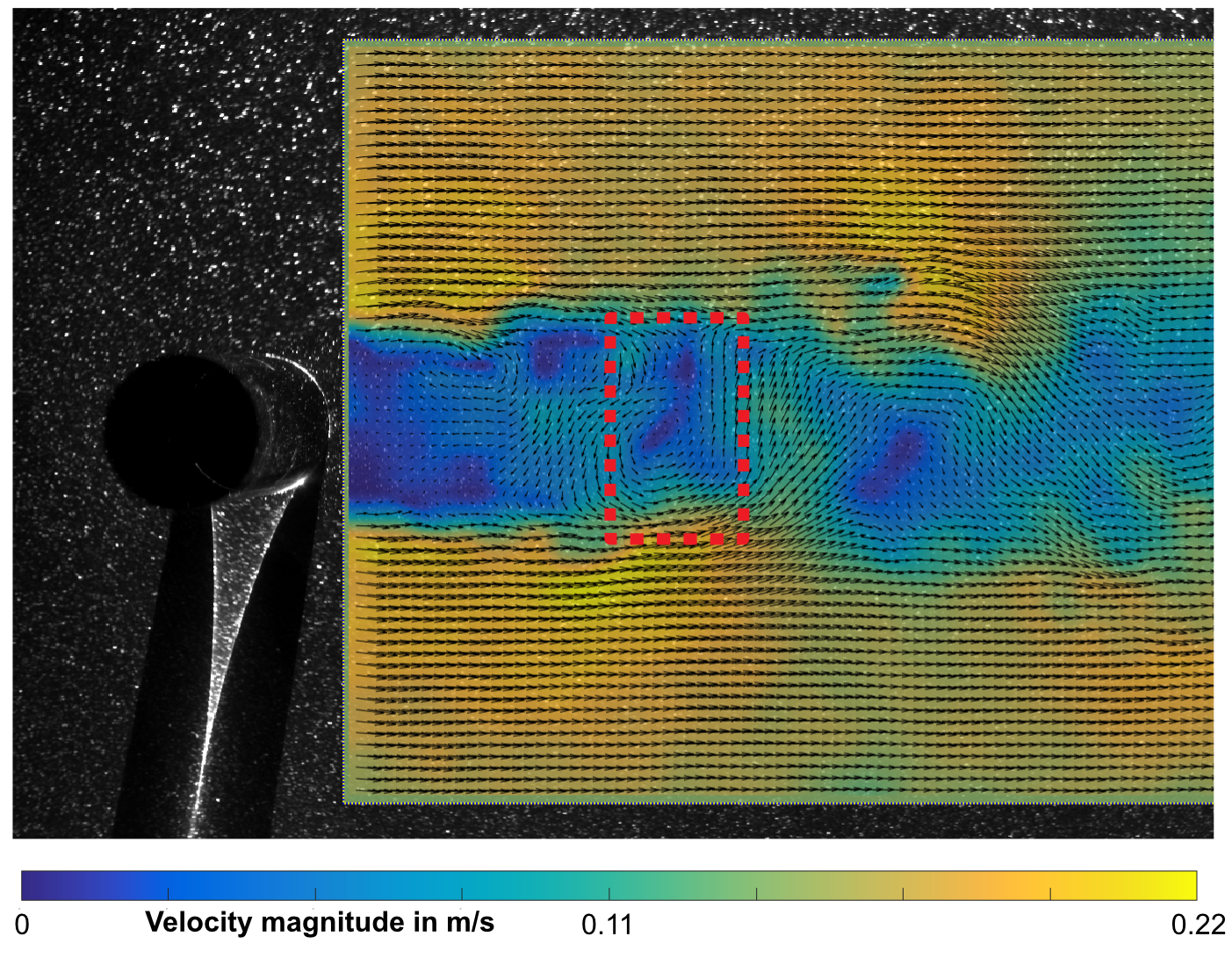}
    \caption{Velocity magnitude behind the cylinder. Shown in red is the position of the box where the velocity component perpendicular to the free stream is extracted.}
    \label{fig:position_box}
\end{figure}
\section{First experiment: measurement of a von-Kármán vortex street and the vortex shedding frequency}
The aim of this experiment is to measure the vortex shedding frequency of a 2D cylinder. The Strouhal number $\mathit{St}$ can be calculated from the shedding frequency $f$, the free stream flow velocity $u_\infty$ and cylinder diameter $D$. The dimensionless number $\mathit{St}$ is given as:
\begin{equation}
\mathit{St} = \frac{f \cdot D}{u_\infty}
\end{equation}
For a cylinder with a smooth surface, this number is constant with $\mathit{St} = 0.21$ for $10^{3} < \mathit{Re} < 10^{4}$. To demonstrate the universal validity of the Strouhal number, a transparent cylinder with a diameter of \SI{15}{\mm} is placed inside the test section. The flow velocity is set to \SI{0.19}{\m/\s}, resulting in a Reynolds number of approximately $2900$. After performing the PIV analysis of 100 image pairs, the instantaneous velocity component perpendicular to the free stream velocity behind the cylinder can be extracted and is shown in Fig. \ref{flow_field_cylinder}. 

\begin{figure}[htb]
    \centering
    \includegraphics[width=0.7\linewidth]{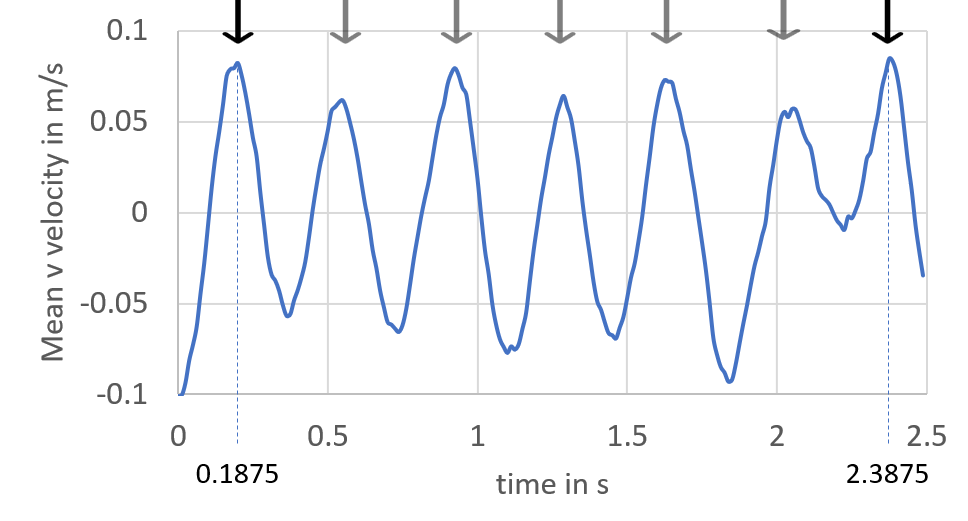}
    \caption{Mean velocity component perpendicular to the free stream velocity in the region behind the cylinder, as shown in Fig.~\ref{fig:position_box}. Detected maxima are indicated by arrows from the top. Deviations from a perfect harmonic oscillation are caused by the non-uniform inflow velocity distribution (see Fig.~\ref{water-tunnel})}
    \label{velocity_fluctuations}
\end{figure}

The shedding frequency is analyzed by selecting a rectangle in PIVlab and exporting the spatially averaged perpendicular velocity component to an Excel sheet for further analysis (Fig.~\ref{fig:position_box}). Due to the clean harmonic nature of the signal, it is sufficient to simply count the maxima of the velocity fluctuations, corresponding to the vortices shed by the cylinder per unit time (Fig.~\ref{velocity_fluctuations}). The current example gives an average frequency of \SI{2.73}{\Hz} with a Strouhal number of $\mathit{St} = 0.214$, which is in excellent agreement with theory. The simple approach of counting maxima of the flow velocity (as indicated by arrows in Fig.~\ref{velocity_fluctuations}) is very instructive to the students and avoids the necessity to perform any further spectral analysis. 

\section{Second experiment: measurement of circulation and lift caused by a thin flat plate}
\begin{figure}[htb]
    \centering
    \includegraphics[width=0.8\linewidth]{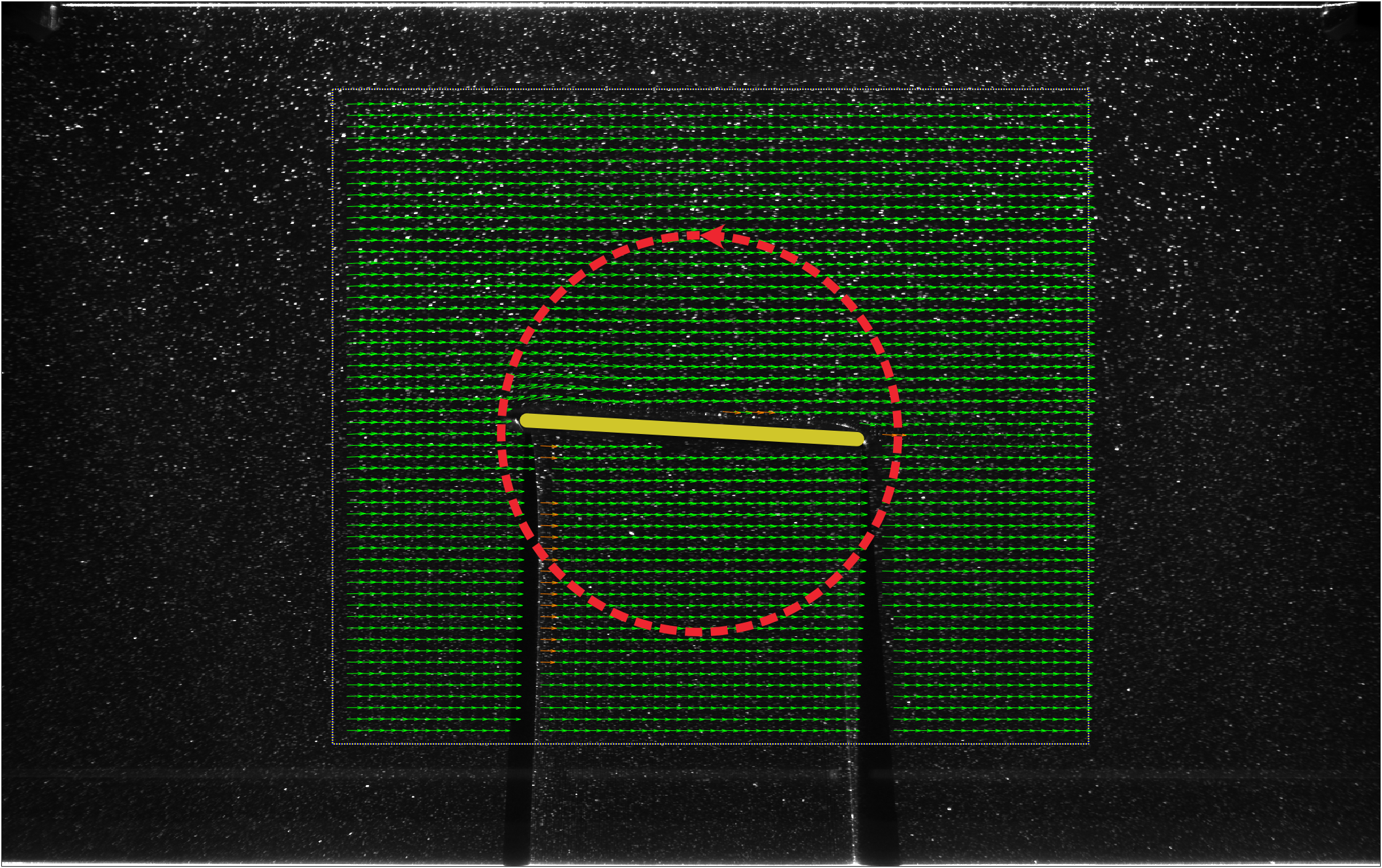}
    \caption{Position of the wing (yellow) and the region used for the area integral as well as for the path integral to calculate $\Gamma$ is shown as a red dashed line. The path integral is calculated in counter clock-wise direction and the average flow direction is indicated by small green vectors from left to right.}
    \label{wing_path}
\end{figure}

\begin{figure}[htb]
    \centering
    \includegraphics[width=0.8\linewidth]{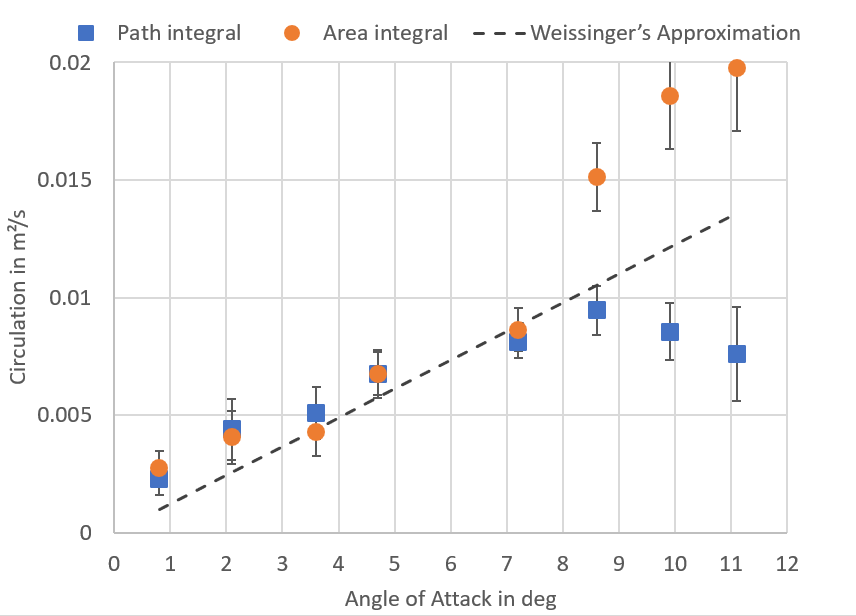}
    \caption{Span-wise circulation derived via path integral of tangential velocity (blue squares), area integral of span-wise vorticity (orange circles), and the Weissinger Approximation (dashed line).}
    \label{results_wing}
\end{figure}

In the second experiment, the circulation of a flat plate with rounded leading and trailing edges (span: \SI{129}{\mm}, chord: \SI{50}{\mm}, thickness: \SI{3}{\mm}) is investigated at eight different angles-of-attack ($0.8^\circ < \alpha < 11.1^\circ )$ and $Re$ = \num{22000}. The measured results are compared to the thin airfoil theory which is calculated based on the Weissinger approximation \cite{weissingerLiftDistributionSweptBack1947}, where the circulation of a thin flat plate is proportional to the angle-of-attack $\alpha$ and given by
\begin{equation}
\Gamma = \pi \cdot u_\infty \cdot c \cdot \alpha \text{,}
\end{equation}
with $\Gamma$ as the span-wise circulation, $u_\infty$ as free stream flow velocity and $c$ as chord length. 
In the following, two different approaches to calculate the circulation from two-dimensional flow fields are compared. As a first approach, the span-wise circulation is derived from the area integral of span-wise vorticity in the vicinity of the wing and, as a second approach, the tangential velocity component is integrated along a path enclosing the wing. The selection of path lines and boundaries is identical for both approaches, as is shown in Fig.~\ref{wing_path}. The time-averaged mean circulation and the standard deviation are calculated for both the area integral and the path integral and compared to theory (Fig.~\ref{results_wing}).

\begin{figure}[htb]
    \centering
    \includegraphics[width=0.8\linewidth]{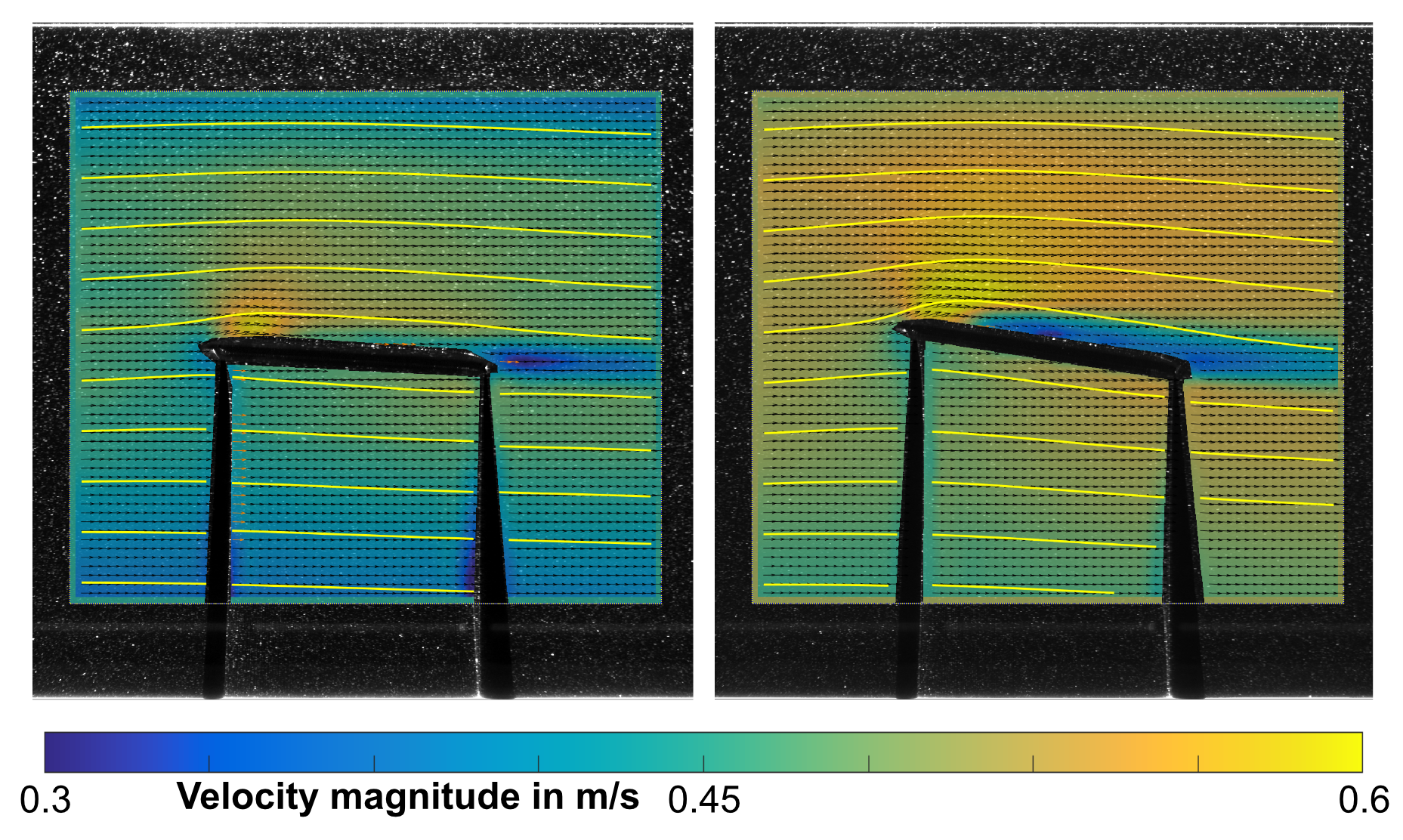}
    \caption{Average velocity fields of the flat plate at an angle-of-attack of $3.6^\circ$ (left) and $8.6^\circ$ (right) at $u_\infty$ = \SI{0.45}{\m/\s}. Streamlines derived from PIV velocity vectors are shown in yellow.}
    \label{wing_velocity_maps}
\end{figure}
\begin{figure}[htb]
    \centering
    \includegraphics[width=0.8\linewidth]{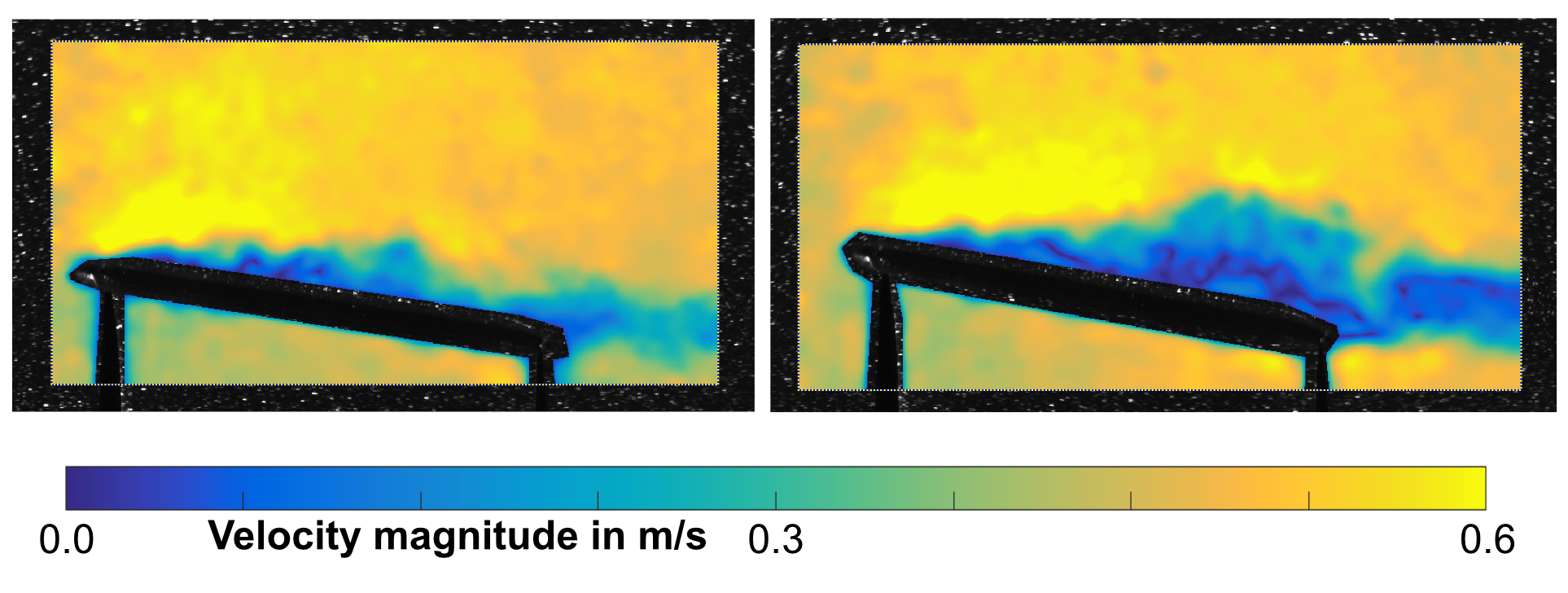}
    \caption{Instantaneous velocity maps of the wing, calculated with wavelet-based optical flow velocimetry (wOFV), yielding slightly better resolved velocity distribution in comparison with interrogation area-based velocimetry. Left: $8.6^\circ$ angle-of-attack ($u_\infty$ = \SI{0.45}{\m/\s}). The flow separation has started, but tends to re-attach at the trailing edge. Right: $9.9^\circ$ angle-of-attack. Flow seperation is clearly visible.}
    \label{wing_velocity_mapswOFV}
\end{figure}

\FloatBarrier

Both methods to calculate the circulation show a good agreement with the theoretical prediction. However, for smaller angles-of-attack ($\alpha \lesssim 8^\circ$), the measured circulation is slightly larger than the prediction. At larger angles-of-attack ($\alpha \gtrsim 8^\circ$) the area integral of the vorticity is larger than the theoretic prediction, while the circulation calculated by a path integral of the tangential velocity is smaller than the circulation predicted by theory. These deviations can be explained by an increasing flow separation with increasing angles-of-attack, as shown by the instantaneous flow fields in Fig.~\ref{wing_velocity_maps} and Fig.~\ref{wing_velocity_mapswOFV}. Large deviations from theory and large standard deviations are to be expected at larger angles-of-attack, because the theoretical model does not take flow separation into account. 

\begin{figure}
    \centering
    \includegraphics[width=0.8\linewidth]{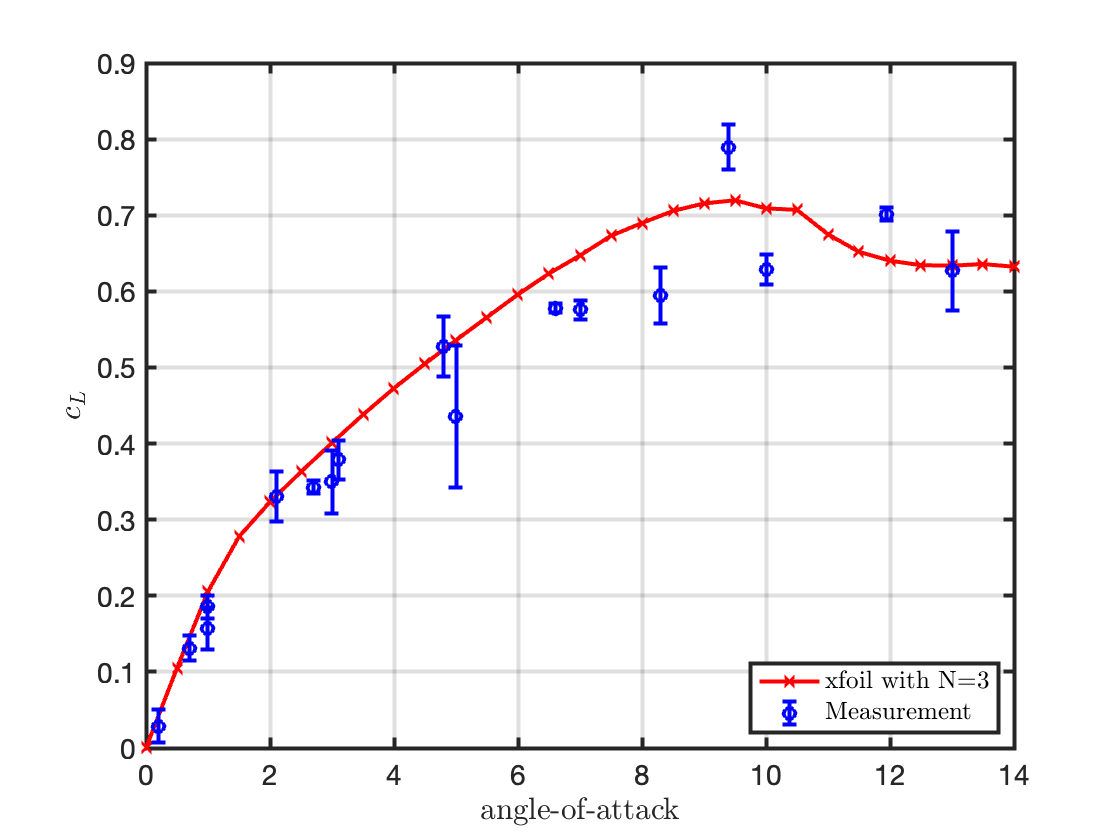}
    \caption{Compaison of the $c_L$ measurement as a function of the angle-of-attack for a NACA-0015 profile with numerical calculations based on XFOIL with a critical $N$-factor of $N=3$}
    \label{Fig:XF-meas}
\end{figure}

\section{Third experiment: measurement of a NACA airfoil and comparison with lift values calculated by XFOIL}

In the third experiment, a NACA-0015 profile, ranging over the complete tunnel width, was tested inside the water tunnel at $\mathit{Re} = \num{19000}$. The circulation was determined by a path integral as described above and then used together with the measured inflow velocity $u_{\infty}$ to calculate the normalized lift coefficient with
\begin{equation}
    c_L = \frac{2 \cdot \Gamma}{u_{\infty} \cdot c} 
\end{equation}

The lift coefficient is shown as a function of the angle-of-attack in Fig. \ref{Fig:XF-meas} in comparison with calculated results from XFOIL, a well established numerical tool based on potential theory \cite{drelaXFOILAnalysisDesign1989}. XFLR5 was used as a graphical user interface for XFOIL, which is helpful as it allows students an easy way to perform numerical calculations \cite{XFLR5}. For transition prediction a critical $N$-factor of $N=3$ was chosen to calculate the location of laminar-turbulent transition inside the boundary layer \cite{schlichtingBoundaryLayerTheory2017}. The experimental results deviate to a certain degree from the numerical prediction, which may be accounted to the limited homogeneity of the flow field (Fig.~\ref{water-tunnel}). However, the general trend of $c_L$ as a function of the angle-of-attack is well captured, which justifies the assumption of a critical $N$-factor of $N=3$ for the described experimental setup.





\section{Conclusion}
Based on the original water tunnel by Prandtl, a modern test facility for teaching purposes was designed. The vertical velocity profile, a von-Kármán vortex street behind a cylinder and the circulation around a flat plate as well as around a NACA profile were measured by PIV. All results show good agreement with theory within the measurement uncertainty. The visual and instructive nature of PIV technology allows the reported findings to be not only measured, but also directly observed by eye. Hence, the developed water tunnel is well suited for the qualitative visualization and quantitative investigation of basic fluid dynamic phenomena. 

The water tunnel has been used successfully to introduce university students to the depth and beauty of fluid dynamics and to teach them principle concepts and perform quantitative measurements. Furthermore, the obtained results show a good agreement with lift theory and simple numerical calculations by XFOIL based on potential theory in combination with the $e^N$-method. Therefore, the experimental setup allows students to directly compare and discuss experimental results with theory and numerical computations.


\section*{Contributions}
Conceptualization: W.T., S.R.; Methodology: W.T., S.R.; Formal analysis and investigation: W.T., S.R.; Writing - original draft preparation: W.T., S.R.; Writing - review and editing: W.T., S.R.; Funding acquisition: S.R.; Resources: W.T., S.R.; Supervision: S.R.
\section*{Acknowledgements}
We are thankful to J. Lemarechal (DLR) for helpful discussions during the construction of the water tunnel. We thank J. Eitelwein, M. Luiking, H. Kraus, T. Weimer, O. Grünberg, A. Grassmann (all FH Kiel) for building a NACA-0015 profile, construction of a laser protection system, performing measurements in the water tunnel and writing a first user manual. We are also thankful to A. P. Schaffarczyk, L. Föhring and J. Kemper (all FH Kiel) for their support in supervising students, review of the article and performing numerical calculations.
\section*{Data availability}
The PIV data can be found at the following \doi{10.5281/zenodo.13982681}.
\bibliographystyle{spphys}       
\bibliography{20241104_MyLibrary.bib}   

\end{document}